\documentstyle[12pt]{article}
\renewcommand{\footnoterule}{\kern -3pt\hrule width 16truecm\kern
2.6pt}

\begin{document}

\centerline{\large Noise-Induced Linearisation and Delinearisation}

\vskip 0.3truecm
\centerline {M I Dykman}
\centerline {Department of Physics, Stanford University,}
\centerline {Stanford, California 94305, USA.}

\vskip 0.3truecm
\centerline {D G Luchinsky\footnote{Permanent address: All Russian
Institute for Metrological Service, Andreevskaya nab 2, 117965 Moscow,
Russia.}}
\centerline {School of Physics and Materials, Lancaster University,}
\centerline {Lancaster, LA1 4YB, UK.}

\vskip 0.3truecm
\centerline {R Mannella}
\centerline {Dipartimento di Fisica, Universit$\grave{\rm a}$ di
Pisa,}
\centerline {Piazza Torricelli 2, 56100 Pisa, Italy.}

\vskip 0.3truecm
\centerline {P V E McClintock, H E Short, N D Stein and N G
Stocks\footnote{Present
address: Department of Engineering, University of Warwick, Coventry,
CV4
7AL, UK.}}
\centerline {School of Physics and Materials, Lancaster University,}
\centerline {Lancaster, LA1 4YB, UK.}

\vskip 0.3truecm
\noindent
\underbar {Abstract}

It is demonstrated, by means of analogue electronic simulation and
theoretically, that external noise can markedly change the character
of the
response of a nonlinear system to a low-frequency periodic field.  In
general, noise of sufficient intensity {\it linearises} the
response.  For
certain parameter ranges in particular cases, however, an increase in
the
noise intensity can sometime have the opposite effect and is shown to
{\it delinearise} the response.  The physical origins of these
contrary
behaviours are discussed.

\newpage
\noindent
INTRODUCTION

\noindent
Noise in physical systems is Janus-like, with two oppositely directed
faces.
Both of them are relevant to an understanding of complex systems$^1$.
The negative, destructive, and therefore ugly, face of noise - which
is
probably the more familiar to most physicists - corresponds to the
familiar
blurring by random fluctuations of otherwise well-defined quantities,
the
randomization of initially ordered systems, and the destruction of
fine
detail
in intricate patterns.  The universality of noise in macroscopic
physical
systems requires that this aspect be taken explicitly into account in
any measurement, because it gives rise to a \lq\lq random error".  The
ugly face of noise is especially obtrusive in relation to studies of
chaotic phenomena$^2$ in real systems, where it is usually the effect
of
internal or external noise that sets the practical lower limit on the
range
of co-ordinate scales over which, for example, fractal effects can
persist.

It is perhaps less well known, but one of the major themes of the
present
volume, that noise can also exhibit a face that is
beautiful, in the sense that its effects can also be positive and
creative.
In stochastic resonance$^{3}$, for example, noise can enhance a weak
periodic signal in a nonlinear system.  It can give rise to
noise-induced
transitions$^{4}$ in which a discontinuous change occurs in the
number of
extrema in a probability density.  Noise can be used to overcome$^{5}$
the phase-locking of ring-laser gyroscopes caused by nonidealities
of their mirrors, thereby linearizing the response. It can create
spatial structures in liquid crystals$^{6}$ and stabilise them in
convecting fluids$^{7}$. Noise also appears to play an important role
in the maintenance of consciousness$^{8}$. These and other examples
are
discussed in earlier chapters of this volume and in the pages that
follow.

It is with the positive, creative, aspect of noise that the present
chapter
is mainly concerned. We consider the effect of external noise on the
passage
of a periodic signal (in principle, of any form) through a nonlinear
system
(in principle, of any form).  For a signal of finite amplitude, on
account
of
the nonlinearity, the output signal $q (t)$ in the absence of noise
will
naturally be distorted compared to the input.  The experimental
observation
that we wish to report and discuss is that this distortion can
apparently
always be removed by the addition at
the input of external white\footnote{For convenience, we will
concentrate
on
the effects of white or quasi-white noise.  Most of the discussion,
however,
is equally applicable qualitatively, if not always quantitatively, to
the
case
of non-white noise, provided that its correlation time is shorter than
the characteristic times of the system under consideration.}
noise of sufficient intensity: a process that might reasonably be
described by the term {\it noise-induced linearisation}.  The
linearized output resulting from this procedure is inevitably noisy
and
so, to focus attention on what happens to the periodic signal itself,
in
what
follows we will discuss how the ensemble average $\langle q (t)
\rangle$
of the output varies with relevant parameters, for example with the
noise
intensity at the input.

These ideas are illustrated by the experimental data shown in Figures
1 and
2, obtained from an analogue electronic circuit model of an overdamped
bistable system (see below).  The input to the model in each case is
the waveform shown at the top, whose frequency is low compared to the
system's reciprocal relaxation time.  For negligible noise intensity
$D$,
the
output response of the system is grossly distorted, as shown by the
upper
$\langle q (t) \rangle$ traces.  As the noise intensity is increased,
however, the distortion steadily diminishes in each case until, in
the
lowest
trace, the output can be seen to be a faithful reproduction of the
input
waveform.  With some exceptions, which will be discussed, this
scenario has
been found experimentally to hold for a very wide range of nonlinear
systems
- monostable as well as bistable, underdamped as well as overdamped,
chaotic
as well as regular - and shapes of the input waveform.

It must be emphasized that the investigations
are still in progress.  Our intention here is to identify
noise-induced
linearisation as an interesting and potentially important
fluctuational
phenomenon that does not seem to have received attention previously
in its
own right, and whose range of occurrence appears to be very wide.  We
make
no claim to a complete understanding of it at this stage, however,
and
would
point out that there are still a number of open questions remaining
to be
resolved.

The plan of the chapter is this.  There follows a general discussion
of the
physical basis of the dramatic results of Figures 1 and 2.  It is
also
pointed
out that the obverse phenomenon of noise-induced {\it delinearisation}
can occur under certain special circumstances.  The third section
presents
a discussion of noise-induced linearisation in relation to the
well-known
and
much-studied case of overdamped motion in a bistable potential; but
it is
argued that, within a certain parameter range, the delinearisation
phenomenon
is also to be anticipated in the same system.  Then there is a
section
describing
investigations of these phenomena in a quite different type of system
- an
underdamped monostable nonlinear oscillator - for which noise-induced
delinearisation, followed by linearisation at higher noise
intensities,
has been both predicted and observed.  Finally, the results are
summarized
and general conclusions are drawn.

\vskip 0.5truecm
\noindent
PHYSICAL BASIS OF NOISE-INDUCED LINEARISATION\newline
\noindent
AND DELINEARISATION

\noindent
The claim for novelty in the results of Figures 1 and 2, and in the
discussion
that follows must be carefully qualified, because the basic idea of
linearisation by added noise will already be
familiar to some through specific applications in certain particular
fields of
science and engineering.  However the proposition that noise-induced
linearisation should exist {\it as a general phenomenon} has not, to
our
knowledge, been enunciated previously.  Nor has it been tested
experimentally,
as we describe below, for a variety of physical systems.

We note that the word {\it linearisation} is commonly used in two
rather
different senses, and that these are exemplified by the results of
Figures
1 and 2.  The fact that a sinusoidal input can pass through the
system
without significant change of shape, as occurs for strong noise in
the
lowest trace
in Figure 1, implies linearity in the sense of a direct
proportionality
between the amplitudes of output and input; this need not
necessarily,
however,
imply that the constant of proportionality must be
frequency-independent.
On the other hand, the results of Figure 2, for a sawtooth waveform
containing not only the fundamental frequency but also its higher
harmonics,
imply the occurrence of linearisation in the \lq\lq Hi-Fi" sense
that the system  becomes non-dispersive within a certain frequency
range
when
the noise intensity is large enough.  It would appear that, for the
types of system considered,\footnote{Obviously, one could try to
devise
special circumstances which might violate this rule.  One possible
example would be a logarithmic amplifier.  However, even though this
would be dispersion-free but highly nonlinear for low noise
intensities, it
seems likely nonetheless that the response would be linearized by
noise,
just
like the other cases that we consider.} linearisation in this latter
sense
automatically implies linearisation in terms of the amplitude as
well, but
not
the converse.

The physical origin of both forms of nonlinearity can readily be
understood,
at least qualitatively, in the following terms.  Where the amplitude
response
of a system to a periodic force is nonlinear, this arises because the
amplitude of the vibrations induced by the force is large compared to
some
characteristic length scale of the system.  The scale in question is
determined by the structure of the region of phase space being
visited by
the
system and by corresponding features in the dynamics.  The effect of
noise
is to smear the system over a larger region of phase space, so that a
variety
of different scales and frequencies then become involved in the
motion,
even
in the absence of periodic driving, and the effective characteristic
scale will usually increase as a result. For sufficiently large noise
intensities, therefore, the amplitude of the force-induced vibrations
will become small compared to the scale (eg small compared to the
average
size of the noise-induced fluctuations), so that the nonlinearity in
the
response amplitude is correspondingly reduced.  Because the system is
then spending an increasing proportion of its time far away from its
attractor(s), at coordinate values where the characteristic time in
responding
to an additional perturbation (the periodic force) will in general be
quite
different and often, in practice, must shorter, than that near the
attractor,
there will be one or more ranges of frequency for which dispersion is
likely
to decrease.  Although the two forms of linearisation arise,
ultimately,
through the same physical processes - the effect of noise in smearing
the system over a larger region of its phase space - there is no
reason to
expect that their onsets will occur at the same noise intensity.

All peaks in spectral densities of fluctuations are broadened by an
increase
in noise intensity, both for overdamped and underdamped systems.
Where the
suppression of dispersion arises (see below) through the broadening
of
a peak centered on zero frequency, one would expect the process to
require a considerably stronger noise intensity than that necessary
to give linearisation in terms of the amplitude of a sinusoidal
waveform at
the fundamental frequency $\Omega$.  This is partly because of the
need
to linearise the response up to larger frequencies, in the case of a
more
complicated waveform of fundamental frequency $\Omega$, in order to
be able
to accommodate the higher Fourier components at multiples of
$\Omega$, and
partly because of the need to obtain a frequency-independent
susceptibility
once linear response has been achieved.

When the noise intensity is sufficient to provide a linear response
in
terms
of amplitudes, the situation can conveniently be discussed in terms of
classical linear response theory$^9$.  Quite generally, the
ensemble-averaged
response of the sytem to a periodic force of frequency $\Omega/2\pi$
can be written in the form

\begin{equation}
\langle q (t) \rangle = \sum\limits_n a (n) [\cos n \Omega t + \phi
(n)]
\end{equation}

\noindent
We ignore here any transient phenomena associated with the initial
application of the stochastic or periodic forces, and suppose that the
system has already reached a stationary state.  We are assuming that
the
noise intensity is large enough (for the given amplitude and frequency
of periodic force) that the response can be characterised by a linear
susceptibility.  The response to a cosine driving force, $A \cos
\Omega t$,
can then be well approximated by

\begin{equation}
\langle q (t) \rangle = a \cos (\Omega t + \phi)
\end{equation}

\noindent
because the higher harmonics in Eq.(1) are negligible, with

\begin{equation}
a \equiv a (1) = A \vert \chi (\Omega) \vert
\end{equation}

\begin{equation}
\phi \equiv \phi (1) = - {\rm arctan} [{\rm Im} \chi (\Omega)/{\rm Re}
\chi (\Omega)]
\end{equation}

\noindent
The susceptibility $\chi (\Omega)$ can be obtained$^{10}$ from the
fluctuation
dissipation theorem$^{9}$.

\begin{equation}
{\rm Re} \chi (\Omega) = \frac {2}{D} {\rm P} \int^{\infty}_{0}
d\omega_1
[\omega^2_1/(\omega_1^2 - \Omega^2)] Q_0 (\omega_1)
\end{equation}

\begin{equation}
{\rm Im} \chi (\Omega) = (\pi \Omega/D) Q_0 (\Omega)
\end{equation}

\noindent
Here, $D$ is the noise intensity, P implies the Cauchy principal
part and $Q_0 (\omega)$ represents the spectral density of
fluctuations of
the system in the absence of the periodic force.  The response to
more
complicated periodic
forces can be considered in a similar way.  Because of the frequency
dependence of $\chi (\Omega)$, the relative amplitudes of Fourier
components at
the harmonics will be modified by the system, so that the response
$\langle q (t) \rangle$ will in general be distorted as compared to
the
input, as observed for weak and intermediate noise intensities in
Figures 1 and 2.

It is also evident from Eqs.(5), (6), however, that where there is a
range (or are ranges) of frequency $\omega$ for which $Q_0 (\omega)$
is
relatively flat, the frequency dependences of $\chi (\Omega)$ and
therefore
of $a, \phi$ will also be correspondingly weak.  Consequently, noise
induced linearisation in the sense of removing dispersion is to be
anticipated whenever an increase of $D$ has the effect of flattening
$Q_0 (\omega)$ within the frequency range of interest.

It is clear from the foregoing discussion that there can be no
generally applicable quantitative theory of noise-induced
linearisation.
The behaviour seen in practice will naturally depend on the system
under
study and, in particular, on the way in which its $Q_0 (\omega)$
evolves
with
noise intensity.  However, the quite general broadening effect
exerted by
noise on the peaks of $Q_0 (\omega)$ implies that there will usually
be at
least one range of frequencies (often centred at zero frequency) for
which
not only linearisation in terms of the amplitudes, but also a
reduction of
dispersion is to be anticipated.

It is important to note, however, that exactly the same physical
processes
which give rise to noise-induced linearisation can also, under special
circumstances, give rise to the opposite effect of noise-induced
delinearisation.  The latter phenomenon is to be anticipated if the
additional
frequencies that become involved (as the result of an increase in
noise
intensity) resonate with the periodic force or with one of its
harmonics,
or if the noise modifies the characteristic reciprocal relaxation time
of the system so that it corresponds to the frequency of the periodic
force.  The standard form of stochastic resonance is related to the
latter
condition.  In such cases, it is to be expected that the promotion of
nonlinearity by noise at intermediate intensities will be followed by
the
more
general phenomenon of noise-induced linearisation at still higher
noise
intensities owing to the usual noise-induced broadening of the
relevant
spectral peak(s).

We now consider, in the following sections two specific examples of
systems whose response can be either linearized or delinearised by
adding
external noise at the input.

\vskip 0.5truecm
\noindent
NOISE-INDUCED LINEARISATION IN AN OVERDAMPED\newline
\noindent
BISTABLE SYSTEM

\noindent
Noise-induced linearisation in terms of the amplitude is already well
known for
the case of overdamped motion in a symmetrical double well potential,
described by

\begin{equation}
\dot q + U^{\prime} (q) = A \cos \Omega t + f (t)
\end{equation}

\begin{equation}
U (q) = - \frac {1}{2} q^2 + \frac {1}{4} q^4
\end{equation}

\noindent
where $A \cos \Omega t$ is the periodic input force and $f(t)$ is
quasi-white zero-mean Gaussian noise of intensity $D$,

\begin{equation}
\langle f (t) f (t^{\prime})\rangle = 2D \delta (t-t^{\prime})
\end{equation}

\noindent
The system described by Eqs.(7)-(9) has been studied intenstively, in
connection particularly with stochastic resonance$^3$, and its
properties
are to a large extent well known and understood.  In particular,
the zero-frequency peak in the spectral density of fluctuations
$Q_0 (\omega)$, caused by interwell transitions, is known to broaden
rapidly
with increasing noise intensity.  Within the range of linear response,
and for $D \ll \Delta U$ where $\Delta U$ is the depth of each
potential
well, which = $\frac {1}{4}$ for the model of Eq (8), the
susceptibility
can$^{11,12}$ be written

\begin{equation}
\chi (\omega) = \sum\limits_n w_n (U^{\prime\prime}_n - i\omega)^{-1}
+
\frac{w_1w_2}{D} (q_1 - q_2)^2 W (W-i\omega)^{-1}
\end{equation}

\noindent
where $w_1 = w_2 = \frac {1}{2}$ are the populations of the two
potential
wells, the $U^{\prime \prime}_n$ are the curvatures of the bottoms of
the
wells at $q_n = (-1)^n$ for $n$ = 1, 2 and $W = W_{12} + W_{21}$ is
the
sum of the interwell transition probabilities $W_{nm}$ from $n \to
m$.  It
is
immediately evident by inspection of Eq (10) that, for $\omega \ll
U^{\prime
\prime}_n$, there will be strong dispersion for weak noise where
$W \stackrel {<}{\sim} \omega$ but negligible dispersion for stronger
noise
when the exponential rise of $W_{nm}$ with $D$ implies $W \gg \omega$.
Figure 3(a)-(c) shows the variation of the response with frequency,
for three different noise intensities.  The full curves correspond
to $\vert \chi^{(1)} \vert^2$; they exhibit a strong dependence on
frequency for weak noise intensity (c), but become much flatter as
the
noise
intensity is increased to a large value (a).  The dashed curves
indicate
the squared ratio of the response at the third harmonic to that at
the fundamental: they provide a quantitative measure of the
dispersion,
which can be seen to decrease rapidly with increasing noise intensity.
Noise-induced linearisation (in the frequency sense) is therefore
indeed
to be
anticipated for the system Eqs.(7)-(9), thereby accounting for the
phenomena
observed in Figures 1 and 2 which were obtained from an analogue
electronic
circuit$^{12}$ of conventional$^{13,14}$ design built to model this
system.

Before going on to consider delinearisation/linearisation phenomena
in a different type of system, we would point out that there is a
parameter range within which the model described by Eqs.(7)-(9) would
also be expected to display noise-induced delinearisation.  To see how
this comes about, note that its response is already known to display
a
giant
nonlinearity$^{15}$ within the parameter range where the noise has
\lq\lq tuned" the reciprocal interwell transition probability in such
a way
that the system makes on average almost exactly one pair of
transitions, at
almost the same phases each time, per cycle of the periodic force.
Clearly,
for very much lower noise intensities, such that the interwell
transitions
are suppressed, the response to {\it weak} periodic forcing about
either
one of the potential wells will be at least approximately linear.
Consequently, there must be a parameter range (low frequency and small
amplitude of the periodic force, very weak noise) for which the system
exhibits some degree of noise-induced delinearisation; but, of
course, it
will
linearize again for larger values of $D$ as already discussed.

\vskip 0.5truecm
\noindent
NOISE-INDUCED DELINEARISATION IN AN UNDERDAMPED\newline
\noindent
MONOSTABLE SYSTEM

\noindent
To demonstrate that noise-induced linearisation/delinearisation
phenomena
are in no way confined to overdamped bistable systems like the one
described by Eqs.(7)-(9), we now show that similar effects are to be
seen in an underdamped, monostable oscillator.  This is an example of
a
case
where delinearisation arises because the effect of the noise is to
modify
the oscillator's eigenfrequency in such a way as to \lq\lq tune" it
through
a harmonic of the periodic force.  The system we consider is described
by the equations

\begin{equation}
\ddot q + 2 \Gamma \dot q + U^{\prime} (q) = A \cos \Omega t + f (t)
\end{equation}

\begin{equation}
U (q) = \frac {1}{2} \omega_0^2 q^2 + \frac{1}{3} \beta q^3 + \frac
{1}{4}
\gamma q^4
\end{equation}

\begin{equation}
\langle f (t) f (t^{\prime})\rangle = 4 \Gamma T \delta (t-t^{\prime})
\end{equation}

\noindent
where $f (t)$ is zero-mean Gaussian white noise of intensity $T$ and
the damping constant is assumed small, $\Gamma \ll \omega_0$.
Provided
that $\gamma >$ 0 and $\vert \beta \vert < 2 \omega_0 \gamma^{\frac
{1}{2}}$,
the potential of Eq (12) has a single minimum and the oscillator is
monostable.
Because of the anharmonic terms in Eq (12), the frequency $\omega (E)$
of the eigenvibrations depends on the oscillator energy $E = \frac
{1}{2}
p^2 + U (q)$, where $p = \dot q$ is the momentum of the oscillator.
For large $E$, the frequency increases with energy, as $E^{\frac
{1}{4}}$.
For small energies, on the other hand,

$$\omega (E) \simeq \omega_0 + \omega^{\prime}_0 E$$

where

$$\omega^{\prime}_0 \equiv \left(\frac {d \omega(E)}{dE}\right)_{E=0}
=
\frac {3}{4} \gamma \omega_0^{-3} - \frac {5}{6} \beta^2
\omega_0^{-5}$$

\noindent
Consequently, for ($\beta^2/\gamma \omega_0^2) > 9/10$, the function
$\omega
(E)$ is nonmonotonic: it decreases with $E$ for small $E$ and then
increases again for larger $E$: $\omega (E)$ is plotted for three
values
of $\beta$ in Figure 4, with $\omega_0 = \gamma = 1$.  Such
nonmonotonicity of $\omega (E)$ in an underdamped oscillator gives
rise to a number of interesting features$^{16-18}$ in its spectral
density of fluctuations that have been reported and discussed
elsewhere.

A situation of particular physical interest arises when the periodic
driving frequency is close to half the eigenfrequency of small
amplitude
vibrations,

\begin{equation}
\vert 2 \Omega - \omega_0 \vert \ll \omega_0
\end{equation}

\noindent
The absorption of a periodic force close to the subharmonic frequency
$\omega_0/2$ can be strongly noise-dependent: in effect, the noise
intensity
can be used to change the energy of the oscillator, and hence adjust
its
eigenfrequency $\omega (E)$ to the second harmonic of the driving
frequency,
giving rise to a nearly resonant absorption.  The \lq\lq tuning"
process is
rather similar to the stochastic resonance that is seen in monostable
systems$^{19}$, but it occurs here for a periodic force applied, not
near
the eigenfrequency, but at approximately half the eigenfrequency.  The
phenomenon is of considerable interest, is of some relevance to
two-photon
absorption processes in optics, and will be discussed  in more detail
elsewhere$^{20}$.  For present purposes, we are mainly concerned with
the
likelihood that the noise-induced resonance will correspondingly give
rise
to a noise-induced nonlinearity of the response.

The motion of the oscillator described by Eqs.(11)-(13) consists of
random vibrations induced by the noise, varying slowly in amplitude
and phase, that get mixed nonlinearly with the periodic vibrations at
$\Omega$
and its overtones induced by the periodic force.  To lowest order in
the
force amplitude, only the vibrations at frequency $\Omega$ are
excited.  If
the noise intensity $T$ is small enough,

\begin{equation}
\langle q (t) \rangle^{(1)} \simeq \frac {A}{3\omega^2} \cos (\Omega
t + \phi^{(1)})
\end{equation}

\begin{equation}
\phi^{(1)} \simeq -2\Gamma/3 \Omega
\end{equation}

\noindent
The existence of the phase shift $\phi^{(1)}$ corresponds to a weak
linear
absorption of energy by the oscillator.  The absorption coefficient
$\kappa$
is defined as the ratio of the energy absorbed per unit time (averaged
over the period) to $A^2$.  In the approximation given by Eqs.(15),
(16),
$\kappa \simeq \kappa^{(1)}$ where

\begin{equation}
\kappa = A^{-2} \overline{\langle \dot q A \cos \Omega t \rangle}
\end{equation}

\begin{equation}
\kappa^{(1)} = \Gamma/9\Omega^2
\end{equation}

\noindent
The resultant vibrations at frequency $\Omega$ give rise in turn to
vibrations at its overtones.  For comparatively small $A$, and when
Eq (14) holds, it is the vibrations at 2 $\Omega \simeq \omega_0$ that
are of primary interest.  The equation of motion for these vibrations
can
be obtained by making the substitution

$$q (t) \simeq \langle q^{(1)} (t) \rangle + q^{(2)} (t)$$

\noindent
in Eq (11) and separating the terms oscillating at frequencies
$\sim 2 \Omega \simeq \omega_0$,

\begin{equation}
\ddot q^{(2)} + 2 \Gamma \dot q^{(2)} + \omega^2_0 q^{(2)} + \beta
[q^{(2)}]^2 + \gamma [q^{(2)}]^3 \simeq f(t) - \frac
{8A^2}{9\omega_0^4}
\beta \cos (2\Omega t + 2 \phi^{(1)})
\end{equation}

\noindent
where we have ignored the renormalisation of the frequency $\omega_0$
by (16/9) $\omega^{\prime}_0 F^2/\omega^2_0$ (which corresponds to a
kind of dynamical Stark shift for a nonlinear oscillator).

Equation (11) constitutes the equation of motion of a
nearly-resonantly-driven
nonlinear oscillator.  Its linear response to the periodic force
given by the final term can be described in a very similar way to the
theory of monostable stochastic resonance$^{19}$, obtaining the
spectral
density $Q_0 (\omega)$ in the absence of the periodic force by the
method developed for the tilted Duffing oscillator$^{16}$, and then
using
the fluctuation dissipation relations (5), (6) (with $T$ in place of
$D$
for
the noise intensity) to determine the susceptibility $\chi
(\Omega)$.  When
this procedure is performed it is found that $\chi (\Omega), a$ and
$\phi$
are strongly dependent on noise intensity, just as anticipated on the
basis
of the simple physical arguments presented above in terms of
stochastic
\lq\lq tuning" of the system.

The theoretical predictions have been tested experimentally on an
analogue electronic model$^{13, 14}$ of Eqs.(11)-(13); the design and
operating procedures will be described in detail elsewhere$^{20}$.
The model was driven by quasi-white noise and by the periodic force
$A \cos
\Omega t$ with its frequency $\Omega$ chosen to be slightly less than
$\omega_0/2$.  With $\beta$ set so as to make $\omega(E)$ nonmonotonic
(Figure 4) one would therefore anticipate a noise-induced resonance
at the
second harmonic for an appropriate value of $D$; one would expect in
turn
that the resultant absorption of energy from the periodic driving
force
would results in a delinearisation of what, in the absence of noise,
had
been a weak and, to a good approximation, linear response at $\Omega$.

For circuit parameters set to $\omega_0$ = 1, $\Gamma$ = 0.011,
$\beta$
= 1.67, $\gamma$ = 1, and with a sinusoidal drive of frequency
$\Omega$ = 0.442 and amplitude $A$ = 0.200, the dependence
on the noise intensity $T$ of the absorption coefficient
$\kappa$ and phase shift $\phi$ of the response at $\Omega$ were
measured
and found to be as shown in Figure 5(a) and (b).  It is immediately
evident
that $\kappa$ and $\phi$ exhibit a maximum and minimum respectively at
$T \simeq$ 0.01.  The occurrence of noise-induced delinearisation was
quantified by computation of the Fourier components of the power
spectrum of the ensemble average $\langle q (t) \rangle$, measured
for different noise intensities.  The ratio $R$ of the intensity of
the
delta
spike at the second harmonic to that at the fundamental frequency
$\Omega$,
providing a measure of the nonlinearity in terms of the amplitude, is
plotted
as a function of $T$ in Figure 6: it can be seen that the
nonlinearity of
the
response rises rapidly with $T$, peaks at $T \simeq$ 0.01 where the
absorption
and phase lag also pass through their maximum values (Figure 5), and
then
decreases again for larger $T$.  Thus, the system (14)-(16) is
exhibiting,
in
turn, noise-induced delinearisation and then noise-induced
linearisation as
$T$ is increased, just as predicted above.

\vskip 0.5truecm
\noindent
CONCLUSION

\noindent
We conclude that the noise-induced linearisation of the response of a
nonlinear
system to a sinusoidal force is a very general phenomenon.
Noise-induced
linearisation in the sense that the frequency dispersion is also
reduced by
noise, so that a non-sinusoidal periodic signal can then pass through
the
system without distortion, also appears to be of wide occurrence: we
suggest
that it is to be anticipated for all hardening potentials provided
that the
fundamental frequency of the force and its relevant harmonics are
very much
lower then the natural frequency of small oscillations in underdamped
systems,
or than the reciprocal relaxation time in the case
of overdamped systems.  An important {\it caveat}, applicable to both
forms of linearisation, is that there may exist a
restricted parameter range within which an increase of the noise
intensity
serves to delinearise the system.  Whether or not a hardening
potential is
a necessary condition remains to be explored: it seems probably that a
hardening potential is required for linearisation in underdamped
systems,
but
possible that any type of binding potential is sufficient in the case
of
overdamped ones. The delinearisation phenomenon is likely to occur in
cases
where the
frequency of the periodic force, or of one of its harmonics, falls
close to
an eigenfrequency (or reciprocal characteristic relaxation time) of
the
system.

Although a preliminary series of electronic analogue experiments on
different
systems has persuaded us that noise-induced linearisation
is widespread, its precise range of occurrence has yet to be
established.  It remains possible, of course, that there are other
physical mechanisms, additional to the one mentioned above, through
which
noise-induced delinearisation can arise.  Further work, both
theoretical
and experimental, will be needed before these effects can be
considered
fully understood and characterised within the larger category of
phenomena
arising through the influence of noise in its positive role as a
creative
force.

\vskip 0.5truecm
\noindent
ACKNOWLEDGEMENTS

\noindent
We gratefully acknowledge support of this work by the Science and
Engineering Research Council (UK), by the Royal Society of London, by
the European Community Directorate General XII, by the Nuffield
Foundation
(London).

\newpage
\noindent
REFERENCES

\begin{itemize}
\item[1.] Nicolis, G., and Prigogine, I. {\it Exploring Complexity: an
Introduction}. New York: Freeman, 1989.

\item[2.] Thompson, J.M.T., and Stewart, H.B. {\it Nonlinear Dynamics
and
Chaos}. New York: Wiley, 1986.

\item[3.] Dykman, M.I., et al \lq\lq Stochastic
Resonance", and references therein.  In this volume, pp xx-yy.

\item[4.] Horsthemke, W., and Lefever, R. {\it Noise-Induced
Transitions}.
Berlin: Springer-Verlag, 1984.

\item[5.] Vogel, K., Risken, H., Schleich, W., James, M., Moss, F.,
Mannella,
R., and McClintock, P.V.E. \lq\lq Colored Noise in the Ring-Laser
Gyroscope: Theory and Simulation." {\it J. Appl. Phys.} {\bf 62}
(1987): 721-723.

\item[6.] Kai, S. \lq\lq Electrohydrodynamic Instability of Nematic
Liquid
Crystals: Growth Process and Influence of Noise." In {\it Noise in
Nonlinear Dynamical Systems}, edited by F. Moss and P.V.E. McClintock,
volume 3, 22-76. Cambridge: Cambridge University Press, 1989.

\item[7.] Deissler, R.J. \lq\lq External Noise and the Origin and
Dynamics
of Structure in Convectively Unstable Systems." {\it J. Stat. Phys.}
{\bf 54} (1989) 1459-1488.

\item[8.] Mandell, A.J., and Selz, K.A. \lq\lq Brain Stem Neuronal
Noise
and
Neocortical \lq Resonance'." {\it J. Stat. Phys.} {\bf 70} (1993):
355-373.

\item[9.] Landau, L.D., and Lifshitz, E.M. {\it Statistical Physics},
3rd Edition. Oxford: Pergamon, 1980.

\item[10.] Dykman, M.I., Mannella, R., McClintock, P.V.E., and Stocks,
N.G. \lq\lq Comment on Stochastic Resonance in Bistable Systems."
{\it Phys. Rev. Lett.} {\bf 65} (1990): 2606.

\item[11.] Dykman, M.I., McClintock, P.V.E., Mannella, R., and Stocks,
N.G. \lq\lq Stochastic Resonance in the Linear and Nonlinear Response
of a Bistable System to a Periodic Field." {\it Sov. Phys. J.E.T.P.
Lett.}
{\bf 52} (1990): 141-144.

\item[12.] Dykman, M.I., Mannella, R., McClintock, P.V.E., and
Stocks, N.G.
\lq\lq Phase Shifts in Stochastic Resonance." {\it Phys. Rev. Lett.}
{\bf 68}, (1992): 2985-2988.

\item[13.] Fronzoni, L. \lq\lq Analogue Stimulations of Stochastic
Processes by Means of Minimum Component Electronic Devices." In {\it
Noise
in Nonlinear Dynamical Systems}, edited by F. Moss and P.V.E.
McClintock,
volume 3, 222-242.  Cambridge: Cambridge University Press, 1989.

\item[14.] McClintock, P.V.E., and Moss, F. \lq\lq Analogue Techniques
for the Study of Problems in Stochastic Nonlinear Dynamics." In {\it
Noise
in Nonlinear Dynamical Systems}, edited by F. Moss and P.V.E.
McClintock,
volume 3, 243-274. Cambridge: Cambridge University Press, 1989.

\item[15.] Dykman, M.I., Mannella, R., McClintock, P.V.E., Stein,
N.D., and
Stocks, N.G. \lq\lq Giant Nonlinearity in the Low Frequency Response
of a
Fluctuating Bistable System." {\it Phys. Rev. E} {\bf 47} (1993):
1629-1632.

\item[16.] Dykman, M.I., Mannella, R., McClintock, P.V.E., Soskin,
S.M.,
and
Stocks, N.G. \lq\lq Noise-Induced Narrowing of Peaks in the Power
Spectra
of Underdamped Nonlinear Oscillators." {\it Phys. Rev. A} {\bf 42}
(1990): 7041-7049.

\item[17.] Soskin, S.M., \lq\lq Evolution of Zero-Dispersion Peaks in
Fluctuation Spectra with Temperature." {\it Physica A} {\bf 180}
(1992):
386-406.

\item[18.] Stocks, N.G., McClintock, P.V.E., and Soskin, S.M.
\lq\lq Observation of Zero-Dispersion Peaks in the Fluctuation
Spectrum of
an
Underdamped Single-Well Oscillator." {\it Europhys. Lett.} {\bf 21}
(1993):
395-400.

\item[19.] Stocks, N.G., Stein, N.D., and McClintock, P.V.E. \lq\lq
Stochastic Resonance in Monostable Systems." {\it J. Phys. A: Math
Gen.}
{\bf 26} (1993): L385-L390.

\item[20.] Dykman, M.I., Luchinsky, D.G., Mannella, R., McClintock,
P.V.E.,
Short, H.E., Stein, N.D., and Stocks, N.G. \lq\lq Noise-Enhanced
Two-Photon
Absorption by a Nonlinear Oscillator", in preparation.
\end{itemize}

\newpage
\noindent
FIGURE CAPTIONS

\begin{itemize}
\item[1.] Noise-induced linearisation for a sinewave passing through
an
electronic model of the overdamped double-well system given by
Eqs.(7)-(9).
The periodic force at the input is shown in the upper trace.  The
ensemble-averaged response $\langle q (t) \rangle$, measured at the
output, is
shown for different noise intensities $D$ in the lower traces.  The
amplitudes
of the latter have been normalised so as to be comparable with the
amplitude of the force, for easier comparison of their relative
shapes.

\item[2.] Noise-induced linearisation for a sawtooth wave
passing through an electronic model of the overdamped bistable system
given
by Eqs.(7)-(9).  The periodic force at the input is shown in the upper
trace. The ensemble-averaged response $\langle q (t) \rangle$,
measured at
the output, is shown for different noise intensities $D$ in the lower
traces.  The amplitudes of the latter have been normalised so as to
be comparable with the amplitude of the force, for easier comparison
of
their
relative shapes.

\item[3.] The normalised response of the system (7)-(9) to a
sinusoidal
driving force, as a function of its frequency $\Omega$, for three
noise
intensities: (a) $D$ = 0.5; (b) $D$ = 0.3; (c) $D$ = 0.02.  The full
curves
represent the squared ratio of the response $a$(1) at the driving
frequency
to the amplitude $A$ of the driving force.  The dashed curves
represent
the squared ratio of the response $a$(3) at the third harmonic of the
driving
frequency to the response $a$(1) at the driving frequency.

\item[4.] Plots of the eigenfrequency $\omega (E)$ of the oscillator
defined by (11)-(13) as a function of the energy in the oscillations,
measured relative to the bottom of the potential well, for (top to
bottom) $\beta$ = 1.30, 1.50 and 1.67, respectively.

\item[5.] A noise induced resonance in an electronic model of the
system
(11)-(13), driven at a frequency slightly less than half the
eigenfrequency
of small amplitude vibrations.  Respectively, (a) the absorption
coefficient
$\kappa$ and (b) the phase shift $\phi$ between the force and the
response at the fundamental are plotted at functions of noise
intensity
$T$.

\item[6.] Noise-induced delinearisation, followed by linearisation,
in an
electronic model of the system (11)-(13), operated under the same
conditions
as for the data of Figure 5.  The ratio $R$ of the delta spike at the
second harmonic to that at the fundamental in the spectral density of
the
ensemble-averaged response $\langle q (t) \rangle$, which provides a
quantitative measure of the linearity of the system, is plotted as a
function of the noise intensity $T$.
\end{itemize}
\end{document}